\documentclass[twocolumn]{aastex62}
\usepackage{graphicx,bm}
\usepackage{color,ulem}
\usepackage{amsmath}

\begin{document}
\title{Potential Impact of Fast Flavor Oscillations on Neutrino-driven Winds and Their Nucleosynthesis}
\shortauthors{Xiong et al.}

\shorttitle{Impact of Fast Flavor Oscillations on Neutrino-driven Winds} 

\author{Zewei Xiong}
\affiliation{School of Physics and Astronomy,
      University of Minnesota, Minneapolis, MN 55455, USA} 
\author{Andre Sieverding}
\affiliation{School of Physics and Astronomy,
      University of Minnesota, Minneapolis, MN 55455, USA} 
\author{Manibrata Sen}
\affiliation{Department of Physics and Astronomy, Northwestern University, Evanston, IL 60208, USA}
\affiliation{Department of Physics, University of California, Berkeley, California 94720, USA}
\author{Yong-Zhong Qian}
\affiliation{School of Physics and Astronomy,
      University of Minnesota, Minneapolis, MN 55455, USA} 
      
\begin{abstract}
    The wind driven by the intense neutrino emission from a protoneutron star (PNS) is 
    an important site for producing nuclei heavier than the Fe group. Because of certain features in 
    the neutrino angular distributions, the so-called fast flavor oscillations may occur very close to the PNS 
    surface, effectively resetting the neutrino luminosities and energy spectra that drive the wind.
    Using the unoscillated neutrino emission characteristics from two core-collapse supernova simulations
    representative of relevant progenitors at the lower and higher mass end, we study the potential effects of 
    fast flavor oscillations on neutrino-driven winds and their nucleosynthesis.
    We find that such oscillations can increase the total mass loss by factors up to $\sim 1.5$--1.7
    and lead to significantly more proton-rich conditions. The latter effect
    can greatly enhance the production of $^{64}$Zn and the so-called light $p$-nuclei $^{74}$Se, $^{78}$Kr, 
    and $^{84}$Sr. Implications for abundances in metal-poor stars, Galactic chemical evolution in general, and
    isotopic anomalies in meteorites are discussed.
\end{abstract}

\section{Introduction}
The collapse of the core of a massive star into a protoneutron star (PNS) results in
profuse emission of neutrinos. Simulations have shown that these neutrinos are likely 
a key ingredient of the explosion mechanisms for core-collapse supernovae (CCSNe)
(see e.g., \citealt{Bethe:1990,Wilson.Mayle.ea:1993,Janka:2012} for reviews). Further, for both the early 
inner ejecta \citep[e.g.,][]{Frohlich.Hauser.ea:2006} and the subsequent long-term neutrino-driven wind 
(NDW) from the PNS \citep[e.g.,][]{Qian.Woosley:1996,Arcones.Janka.ea:2007}, neutrino interactions with the 
material determine the conditions governing the associated nucleosynthesis, especially the electron 
fraction or neutron-to-proton ratio. Such neutrino-heated ejecta is potentially a major site
for producing $^{64}$Zn and the light $p$-nuclei $^{74}$Se, $^{78}$Kr, $^{84}$Sr, and
$^{92}$Mo \citep[e.g.,][]{Hoffman.Woosley.ea:1996}. The latter nuclei
are so named because they cannot be accounted for by neutron-capture processes. 
Similar neutrino-driven outflows are also expected from
neutron star mergers (NSMs) \citep[e.g.,][]{Just.Bauswein.ea:2015,Perego.Rosswog.ea:2014}.  

Experiments with solar, atmospheric, reactor, and accelerator neutrinos have established that
neutrinos oscillate among different flavors (see e.g., \citealt{PhysRevD.98.030001} for a review). 
Whereas the effects of neutrino oscillations on CCSN explosion and nucleosynthesis have been
studied for a long time \citep[e.g.,][]{1992ApJ...389..517F,1993PhRvL..71.1965Q}, such effects are
yet to be included self-consistently in CCSN simulations. A proper treatment is difficult partly
because the regular neutrino transport changes from diffusion inside the PNS to free-streaming 
outside it \citep[e.g.,][]{Richers.McLaughlin.ea:2019}. In addition, because forward scattering 
among neutrinos causes highly nonlinear flavor evolution of the dense neutrino gas, following
such evolution is effectively a new type of neutrino transport in the flavor space (see e.g.,
\citealt{Duan:2010bg,Mirizzi:2015eza} for reviews).

In this paper, we consider an interesting scenario originally suggested by \cite{Sawyer:2009}, 
where fast oscillations in a dense neutrino gas with certain angular 
distributions can cause neutrinos of different flavors with 
different initial emission characteristics to quickly approach flavor equilibration.
Were such fast oscillations to occur near the surface of a PNS, they would reset the effective 
luminosities and spectra for neutrino interactions outside the PNS. In addition, were
flavor equilibration to be obtained, there would be no need to consider further flavor evolution.
Assuming the above features of fast flavor oscillations, we explore their impact on the NDWs and the
associated nucleosynthesis. We find that such oscillations can increase the total mass loss by 
factors up to $\sim 1.5$--1.7 and lead to significantly more proton-rich conditions. The latter effect 
can greatly enhance the production of $^{64}$Zn and the light $p$-nuclei 
$^{74}$Se, $^{78}$Kr, and $^{84}$Sr. 

This paper is organized as follows.
We describe our NDW models with fast flavor oscillations in \S\ref{sec:setup}.
We present the effects of such oscillations on NDWs in \S\ref{sec:dynamics} and those
on nucleosynthesis in \S\ref{sec:nucleo}. Implications for abundances in metal-poor stars, 
Galactic chemical evolution in general, and isotopic anomalies in meteorites
are discussed in \S\ref{sec:implications}.
Further discussion and conclusions are given in \S\ref{sec:conclusions}.

\section{Modelling NDWs with Fast Flavor Oscillations}
\label{sec:setup}

Fast neutrino flavor oscillations have been examined by many studies
\citep[e.g.,][]{Izaguirre:2016gsx,Dasgupta:2017oko,Capozzi:2017gqd,Abbar:2017pkh,Yi:2019hrp,Capozzi.Dasgupta.ea:2019,Martin:2019gxb,Abbar.Volpe.ea:2019,azari2019linear,nagakura2019fast,fast2020abbar,glas2019fast,azari2020fast}.
\citet{Wu.Tamborra:2017a} found that conditions for such oscillations appear to be
ubiquitous in NSMs, which has important implications for nucleosynthesis and future kilonova
observations if flavor equilibration is obtained \citep{Wu.Tamborra.ea:2017b}. 
Whether conditions for fast flavor oscillations are prevalent in CCSNe depends sensitively on 
detailed neutrino transport calculations. We do not take up this important issue here.
Instead, we assume that such oscillations occur during the long-term cooling of the PNS
and explore their effects on the NDW and the associated nucleosynthesis.

We consider three different assumptions for neutrino oscillations.  
We take the neutrino luminosities and energy spectra from representative CCSN simulations 
and assume no flavor oscillations for our baseline models. As a limiting case, we
assume that fast flavor oscillations result in complete flavor equilibration for the neutrino
fluxes driving the winds. Specifically, the oscillated spectral fluxes are
\begin{subequations}
    \begin{alignat}{2}
    F_{\nu_e}^\mathrm{osci} = F_{\nu_x}^\mathrm{osci} &= \frac{F_{\nu_e}^0+ 2 F_{\nu_x}^0}{3} ,\\
    F_{\bar\nu_e}^\mathrm{osci} = F_{\bar\nu_x}^\mathrm{osci} &= \frac{F_{\bar\nu_e}^0 + 2 F_{\bar\nu_x}^0}{3},
    \label{eq:complete_flavor_equilibrium_spectra}
    \end{alignat}
\end{subequations}
where $F_{\nu_e}^0$, $F_{\bar\nu_e}^0$, and $F_{\nu_x}^0=F_{\bar\nu_x}^0$ are
the unoscillated spectral fluxes and $x$ stands for $\mu$ or $\tau$. Note, however, that
the above expressions represent an approximate limit as flavor equilibration is subject
to conservation of flavor lepton number \citep{Dasgupta:2017oko,Abbar.Volpe.ea:2019}. 
As an intermediate case, we take the results from
\cite{Dasgupta:2016dbv}, who modelled fast flavor oscillations by introducing small artificial
anisotropies in the neutrino angular distributions from a spherically symmetric CCSN simulation.
In this case, the oscillated spectral fluxes are
\begin{subequations}
    \begin{alignat}{4}
    F_{\nu_e}^\mathrm{osci} &= P_{ee} F_{\nu_e}^0 + (1-P_{ee}) F_{\nu_x}^0 ,\\
    F_{\bar\nu_e}^\mathrm{osci} &= P_{\bar e\bar e} F_{\bar\nu_e}^0 + (1-P_{\bar e\bar e}) F_{\bar\nu_x}^0 ,\\
    F_{\nu_x}^\mathrm{osci} &= \frac{1-P_{ee}}{2} F_{\nu_e}^0 + \frac{1+P_{ee}}{2} F_{\nu_x}^0 ,\\
    F_{\bar\nu_x}^\mathrm{osci} &= \frac{1-P_{\bar e\bar e}}{2} F_{\bar\nu_e}^0 + \frac{1+P_{\bar e\bar e}}{2} F_{\bar\nu_x}^0 ,
    \end{alignat}
\end{subequations}
where $P_{ee}=0.68$ and $P_{\bar e \bar e}=0.55$. Note that these values are not based on
a self-consistent CCSN simulations but represent plausible results intermediate between the cases of no flavor
oscillations and complete flavor equilibration. There is no need to consider further neutrino oscillations
for the case of complete flavor equilibration. We do not consider other types of oscillations for the 
case of intermediate flavor conversion, either.

\begin{figure}[!ht]
    \centering
    \includegraphics[width=\linewidth]{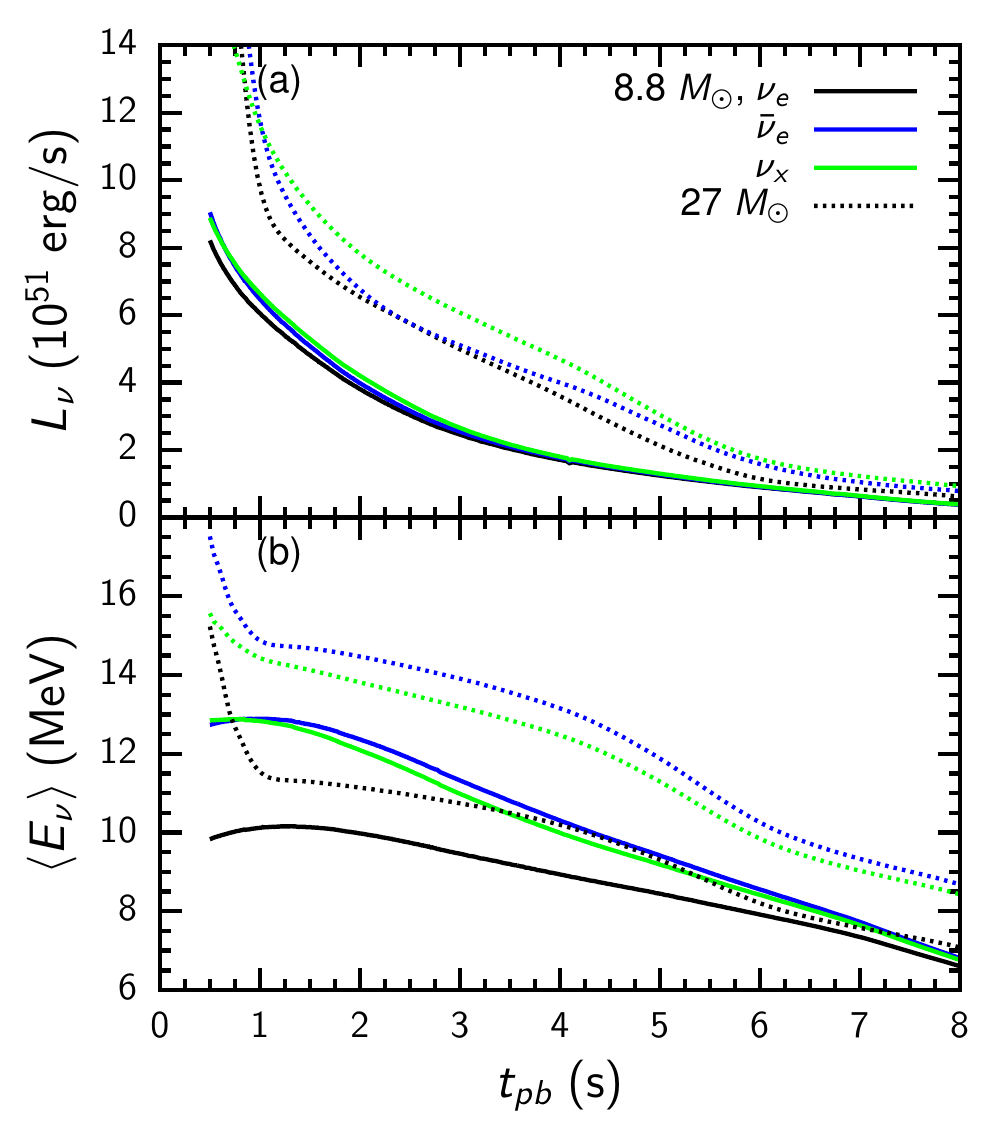}
    \caption{Time evolution of (a) luminosities $L_\nu$ and (b) average energies $\langle E_\nu \rangle$ 
    for $\nu_e$ (black), $\bar\nu_e$ (blue), and $\nu_x$ (green) for models e8.8 (solid curves) 
    and s27 (dashed curves).}
    \label{fig:nu_spec}
\end{figure}

For the unoscillated spectral fluxes, we take the neutrino luminosities and average energies as functions 
of time from two CCSN simulations. The normalized spectrum is taken to be Fermi-Dirac with zero chemical 
potential and is characterized by a temperature for each neutrino species.
Neutrino emission is not expected to vary drastically among progenitors whose CCSNe would result in
long-term NDWs. Nevertheless, there are some trends and characteristics that distinguish the emission for
relevant progenitors \citep{Janka:2012}. In particular, more massive 
progenitors are expected to be associated with an extended phase of accretion-driven neutrino emission,
whereas low-mass ones can explode more promptly and thus exhibit a faster decline of the neutrino luminosities.  
We adopt the results from two representative state-of-the-art CCSN simulations
in spherical symmetry with detailed long-term neutrino transport. One simulation
\citep{Huedepohl.Mueller.ea:2010} was based on the $8.8\,M_\odot$ progenitor model of \citet{Nomoto:1984}.  
We will refer to the corresponding neutrino emission model as model e8.8. The other model, reported in 
\citet{Mirizzi:2015eza} and referred to as model s27, is from a simulation based on a $27\,M_\odot$ progenitor model 
of \citet{Woosley.Heger.ea:2002}. Note that explosion was triggered artificially for model s27.
The time evolution of neutrino luminosities and average energies for models e8.8 and s27 is shown
in Fig.~\ref{fig:nu_spec}. These results are considered as representative of
neutrino emission for relevant progenitors at the lower and higher mass end.

All neutrinos are assumed to be emitted from a sharp neutrino sphere whose radius $R_\nu$ evolves with
time post core bounce $t_\mathrm{pb}$ as
\begin{equation}
    R_\nu=\left[ 12 +8.5~ \left(\frac{1\,\mathrm{s}}{t_\mathrm{pb}}\right)^{0.7} \right] \, \mathrm{km}.
    \label{eq:rnu}
\end{equation}
The above result is fitted to model e8.8 and assumed to apply to model s27 as well.
As the timescales for evolution of $R_\nu$ and neutrino emission are sufficiently long, we
model the NDW as a steady-state outflow from the neutrino sphere. For a specific $t_{\rm pb}$,
we solve the velocity $v$, density $\rho$, temperature $T$, and electron fraction $Y_e$ of the NDW
as functions of radius $r$ from the following equations:
\begin{subequations}
\label{eq:wind_equations}
    \begin{alignat}{2}
        \dot{M} &= 4\pi r^2 \rho v ,\\
        v \frac{d v}{d r} &= -\frac{1}{\rho} \frac{d P}{d r} - \frac{G M_\text{PNS}}{r^2},\\
        \frac{d \epsilon}{d r} &=
        \frac{P}{\rho^2} \frac{d \rho}{d r}
        + \frac{\dot{q}}{v} ,\\
        \frac{d Y_e}{d r} &=\frac{1}{v}[(\lambda_{\nu_e n}+\lambda_{e^+ n})Y_n
        -(\lambda_{\bar{\nu}_e p}+\lambda_{e^- p})Y_p],
    \end{alignat}
\end{subequations}
where $\dot M$ is the mass loss rate, $P$ is the pressure,
$G$ is the gravitational constant, $M_{\rm PNS}=1.282\,M_\odot$ is the PNS mass,
$\epsilon$ is the internal energy per unit
mass, $\dot{q}$ is the rate of net energy gain per unit mass, $Y_n$ ($Y_p$) is
the number of free neutrons (protons) per baryon, i.e., the number fraction of
free neutrons (protons), and $\lambda_{\nu_e n}$ ($\lambda_{e^- p}$) and
$\lambda_{\bar{\nu}_e p}$ ($\lambda_{e^+ n}$) are the rates per target nucleon
for the $\nu_e$ and $\bar\nu_e$ absorption reactions (inverse reactions),
respectively. Because neutrino heating mainly occurs before any significant 
synthesis of heavy elements, the nuclear composition is taken to be free nucleons 
in nuclear statistical equilibrium (NSE) with $\alpha$-particles when the above
equations are solved. The details of $P$, $\epsilon$, $\dot q$, and the rates for
determining $Y_e$ are given in \cite{Xiong.Wu.ea:2019}. For each of the three
cases of neutrino oscillations, the corresponding
neutrino rates for heating the wind and determining $Y_e$ are used.

At each $t_\mathrm{pb}$, the inner boundary conditions of the steady-state solution 
are specified by $\dot q=0$, $dY_e/dr=0$, and $T=T_{\nu_e}$ at $r=R_\nu$, where
$T_{\nu_e}$ is the temperature of the initial $\nu_e$ spectrum.
These conditions correspond to approximate equilibrium between matter and neutrinos
at the neutrino sphere. In principle, the mass loss rate $\dot M$ should be obtained 
as an eigenvalue to satisfy the outer boundary condition imposed by the earlier CCSN
ejecta on the NDW. Because the outer boundary, or the wind termination point, 
at $r=r_{\rm wt}$ is outside the region of efficient neutrino heating, the actual 
solution is well approximated by the transonic solution for $r\lesssim r_{\rm wt}$ \citep{Qian.Woosley:1996}. 
As the outer boundary evolves with $t_{\rm pb}$ and varies with CCSN progenitors,
for simplicity, we adopt the transonic solution up to $r_{\rm wt}=1000$ km for both models
e8.8 and s27 \citep[e.g.,][]{Arcones.Janka.ea:2011,Arcones.Janka.ea:2007}, but also 
consider $r_{\rm wt}=500$ km for model s27 to cover the range of NDW solutions
\citep[e.g.,][]{Wanajo.Janka.ea:2011}. At $r_{\rm wt}=1000$ km, the temperature is 
typically below $1\,\mathrm{GK}$, which corresponds to NDW 
interaction with the CCSN ejecta without generating a strong shock.
At $r_{\rm wt}=500$ km, the temperature can be up to $3\,\mathrm{GK}$, 
which is still appropriate for post-shock material.

The steady-state profiles are calculated for many values of $t_{\rm pb}$ to give
$v$, $T$, $\rho$, and $Y_e$ on a 2D grid of $r$ and $t$. The trajectory of a tracer 
in the NDW is calculated from $r(t)=r(t_0)+\int_{t_0}^tv(r(t'),t')dt'$ and the
associated $T$, $\rho$, and $Y_e$ are obtained by interpolating the corresponding
profiles on the 2D grid. As the steady-state solutions have approximately reached 
the asymptotic velocity at $r_{\rm wt}=1000$ km, we assume $v(t)=v_{\rm wt}$ and
\begin{equation}
r(t)=r_{\rm wt}+v_{\rm wt}(t-t_{\rm wt})
\end{equation}
for $r>1000$ km. Here and below, the subscript ``wt'' denotes quantities at 
$r=r_{\rm wt}$. For $r_{\rm wt}=500$ km, the steady-state NDWs should 
be slowed down before reaching the asymptotic velocity. So we assume
\begin{equation}
r(t)=r_{\rm wt}\left[1-\frac{v_{\rm wt}t_{\rm wt}}{r_{\rm wt}}+
\frac{v_{\rm wt}t_{\rm wt}}{r_{\rm wt}}\left(\frac{t}{t_{\rm wt}}\right)^3\right]^{1/3}    
\end{equation}
for $r>r_{\rm wt}$ in this case, where $v(t)=dr/dt$ has an asymptotic value of 
$v_{\rm wt}[r_{\rm wt}/(v_{\rm wt}t_{\rm wt})]^{2/3}<v_{\rm wt}$ \citep{Wanajo.Janka.ea:2011}.
For both $r_{\rm wt}=1000$ and 500 km, we assume $r^2\rho v=r_{\rm wt}^2\rho_{\rm wt}v_{\rm wt}$ 
and $T^3/\rho=T_{\rm wt}^3/\rho_{\rm wt}$ to extrapolate $\rho(t)$ and $T(t)$ for $r>r_{\rm wt}$
\citep[e.g.,][]{Ning.Qian.ea:2007,Wanajo.Janka.ea:2011}. In this approach, we neglect any entropy 
jump due to the possible formation of a reverse shock at $r_{\rm wt}$ \citep{Arcones.Janka.ea:2007}.
The evolution of $Y_e$ is followed along with nucleosynthesis for the tracer and will be discussed 
in \S\ref{sec:nucleo}.

\begin{figure}
    \centering
    \includegraphics[width=\linewidth]{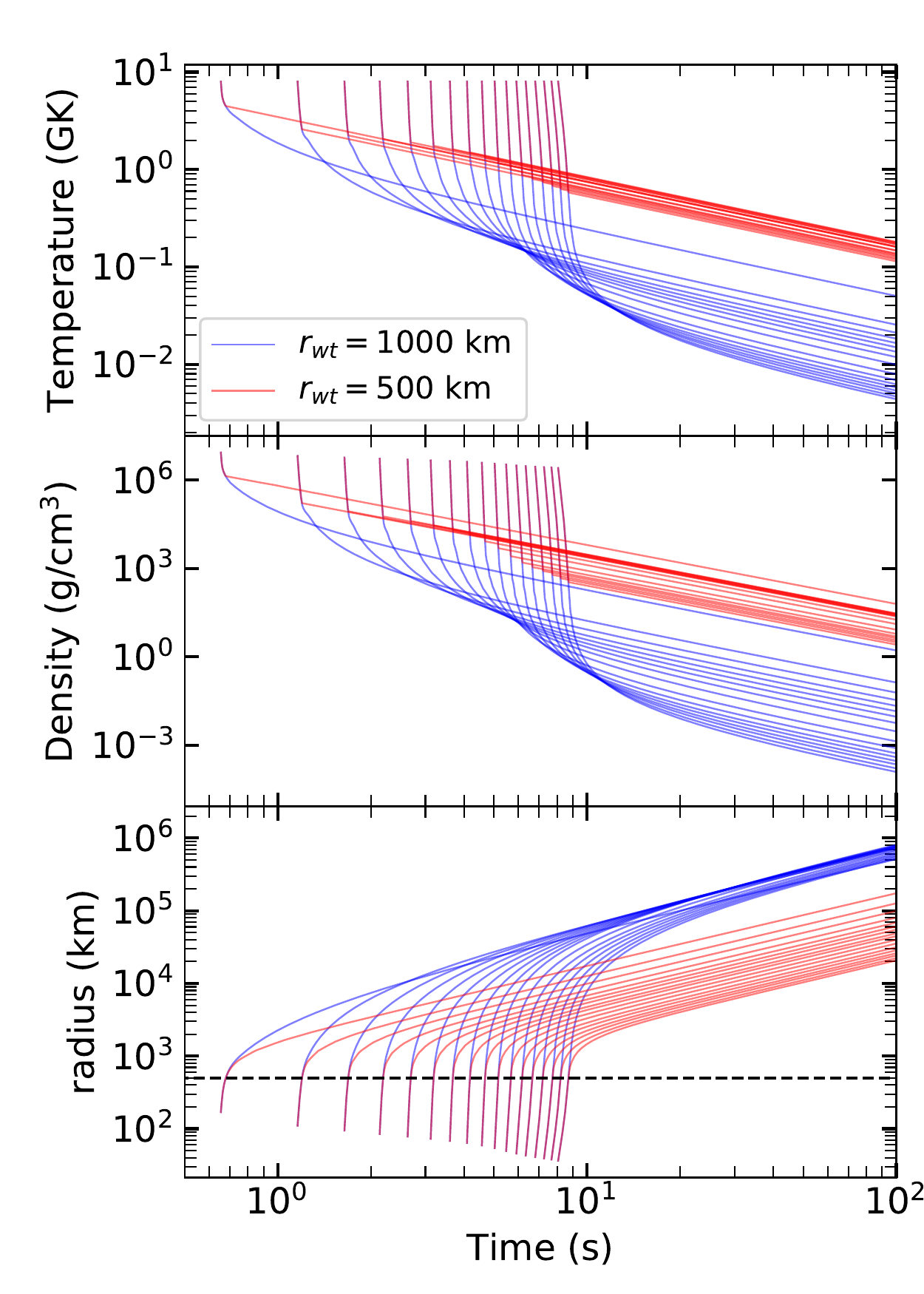}
	\caption{Evolution of temperature (top panel), density (middle panel), and radius (bottom panel) 
	for example tracers in model s27 with $r_{\mathrm{wt}}=1000$ (blue curves) and 
	500 km (red curves) in the case of no flavor oscillations. The horizontal dashed line in the
	bottom panel indicates $r_{\mathrm{wt}}=500$~km.}
    \label{fig:traj_examples}
\end{figure}

Figure \ref{fig:traj_examples} shows $T(t)$, $\rho(t)$, and $r(t)$ for example tracers in model s27 with 
$r_{\mathrm{wt}}=1000$ and 500 km in the case of no flavor oscillations. 
It can be seen that for $r_{\mathrm{wt}}=500$ km, the temperature and density stay at relatively high 
values for longer time. The corresponding effects on nucleosynthesis will be discussed 
in \S\ref{sec:nucleo}.

\section{Effects of Fast Flavor Oscillations on NDWs}
\label{sec:dynamics}
\begin{figure}[!ht]
    \centering
    \includegraphics[width=\linewidth]{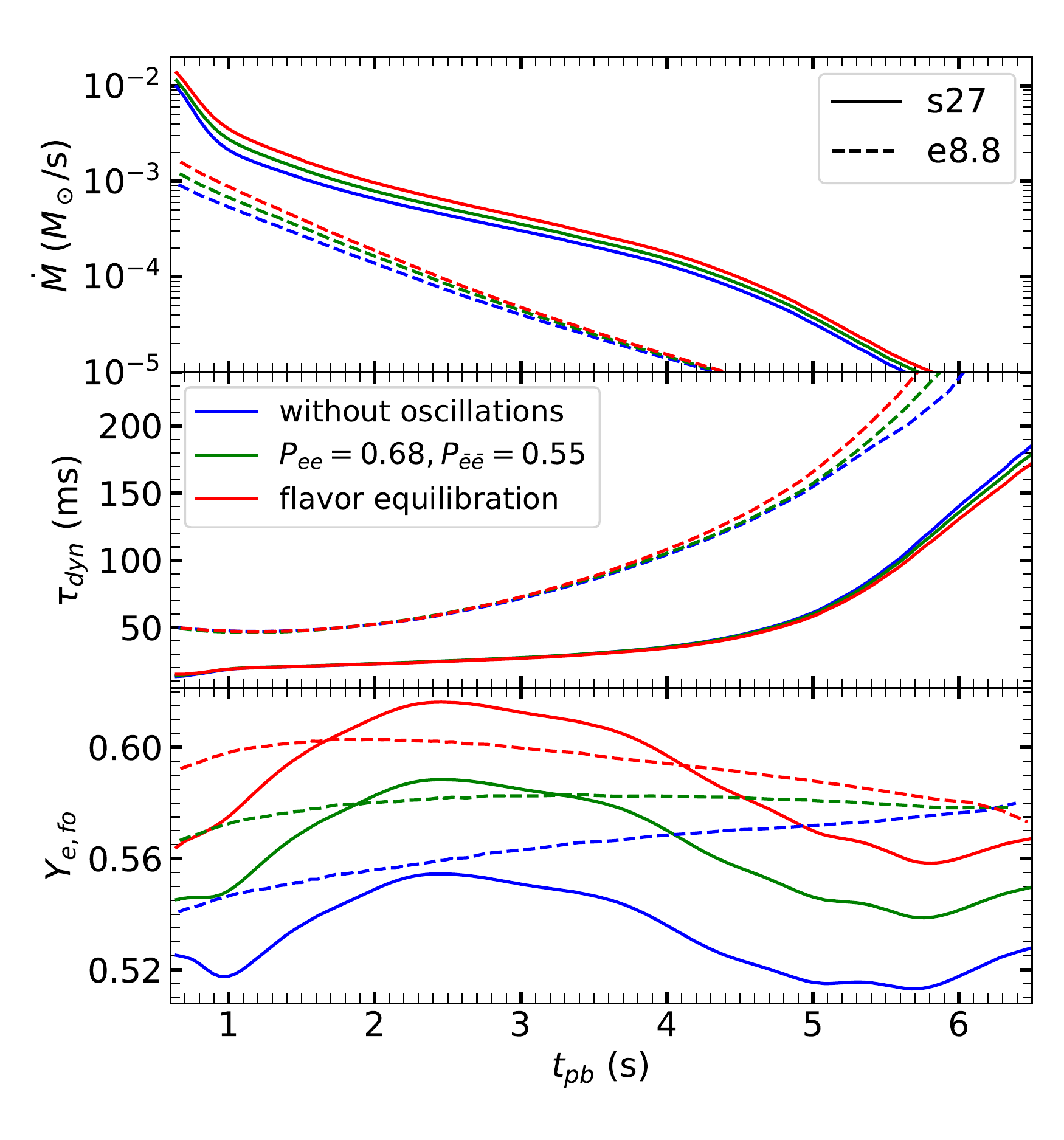}
    \caption{Evolution of the mass loss rate $\dot{M}$ (top panel),
    dynamical timescale $\tau_{\rm dyn}$ (middle panel), and
    freeze-out electron fraction $Y_{e,{\rm fo}}$ for the NDW.
	Each panel shows the results for models e8.8 (dashed curves)
	and s27 (solid curves), both with $r_{\rm wt}=1000$ km, in the
	cases of no flavor oscillations (blue curves), intermediate 
	flavor conversion (green curves), and complete flavor
	equilibration (red curves).}
    \label{fig:dynamics}
\end{figure}

The main heating processes for driving the wind are
\begin{subequations}
\label{eq:nupn}
    \begin{alignat}{2}
  \nu_e+n&\to p+e^-,\\
  \bar\nu_e+p&\to n+e^+,
    \end{alignat}
\end{subequations}
which also set the important parameter $Y_e$ for nucleosynthesis. 
Figure~\ref{fig:nu_spec} shows that nearly for the entire duration of
the NDW, $\nu_x$ ($\bar\nu_x$) has higher luminosity and average
energy than $\nu_e$ ($\bar\nu_e$) for model s27. In comparison,
for model e8.8, the luminosities are close for all neutrino species,
$\bar\nu_x$ and $\bar\nu_e$ also have very similar average energy,
but $\nu_x$ still has higher average energy than $\nu_e$.
As fast flavor oscillations convert $\nu_x$ ($\bar\nu_x$) into $\nu_e$ ($\bar\nu_e$),
heating by the reactions in Eq.~(\ref{eq:nupn}) is enhanced, which produces a higher
mass loss rate $\dot M$ (see Fig.~\ref{fig:dynamics}), and hence, a higher 
net mass loss in the NDW. Indeed, as shown in Table~\ref{tab:wind_properties}, 
the net mass loss increases by a factor of $\sim 1.7$ and 1.5 (1.3 and 1.2) for 
models e8.8 and s27, respectively, in the case of complete flavor equilibration 
(intermediate flavor conversion). The mass loss rate for model s27 is higher than
that for model e8.8 because of the higher neutrino luminosities and average energies 
for the former.

\begin{table*}[t]
    \centering
    \caption{Effects of fast flavor oscillations on the NDW and
	the $\nu p$ process.}
    \begin{tabular}{ccccc} \hline \hline
        Models & Oscillations &  Mass loss $(10^{-3} M_\odot)$ &   $Y_{e,\mathrm{fo}}^\mathrm{max}$  & $\Delta_{n,\mathrm{max}}$ \\ \hline
         e8.8, $r_{\rm wt}=1000$ km    &  without oscillations        &  0.79  & 0.64  & 3.12 \\
         e8.8, $r_{\rm wt}=1000$ km    &   $P_{ee}=0.68$,  $P_{\bar e \bar e}=0.55$  & 1.06  & 0.65 & 3.70 \\
         e8.8, $r_{\rm wt}=1000$ km    &  flavor equilibration &   1.33  &  0.66 & 3.91 \\ \hline
         s27, $r_{\rm wt}=1000$ km     &  without oscillations         &  3.64  & 0.55 & 1.10 \\
         s27, $r_{\rm wt}=1000$ km     &    $P_{ee}=0.68$,  $P_{\bar e \bar e}=0.55$  & 4.43 & 0.59 & 2.31 \\
         s27, $r_{\rm wt}=1000$ km     &   flavor equilibration &  5.48  &  0.61 & 3.67 \\ \hline
         s27, $r_{\rm wt}=500$ km     &   without oscillations         &  3.64  & 0.55 & 2.48 \\
         s27, $r_{\rm wt}=500$ km     &    $P_{ee}=0.68$,  $P_{\bar e \bar e}=0.55$ &  4.43 & 0.59 & 5.34 \\
         s27, $r_{\rm wt}=500$ km     &   flavor equilibration &    5.48  &  0.61 & 8.94 \\ \hline
    \end{tabular}
    
    \label{tab:wind_properties}
\end{table*}

As we will discuss in \S\ref{sec:nucleo}, production of heavy nuclei in the NDW
depends on the ratio of protons to seed nuclei (heavier than $^4$He), which is
sensitive to the dynamical timescale for the earlier expansion of a tracer.
Following \citet{Qian.Woosley:1996}, we define this quantity as
\begin{equation}
    \tau_{\rm dyn}=(r/v)_{T=0.5\,\mathrm{MeV}},
    \label{eq:taudyn}
\end{equation}
which is calculated when the temperature of a tracer drops to
$T=0.5\,\mathrm{MeV}$. Higher neutrino luminosities and average
energies drive faster winds. So $\tau_{\mathrm{dyn}}$ for model 
e8.8 is longer than that for model s27 for the same $r_{\rm wt}=1000$ km,
and $\tau_{\mathrm{dyn}}$ for both models increases with time
(see Fig.~\ref{fig:dynamics}). On the other hand,
fast flavor oscillations affect $\tau_{\mathrm{dyn}}$ for both models 
only at late times when there is little mass ejected in the NDW 
(see Fig.~\ref{fig:dynamics}).
They also do not significantly affect the entropy (not shown), which is
another important quantity for setting the proton-to-seed ratio.

Yet another important quantity for nucleosynthesis in the NDW is  
the freeze-out electron fraction $Y_{e,{\rm fo}}$
(taken at $T=3$~GK), which is determined by the competition between the reactions in Eq.~(\ref{eq:nupn}). 
Specifically, $Y_{e,{\rm fo}}>0.5$ when $\langle E_{\bar\nu_e}\rangle-\langle E_{\nu_e}\rangle$ lies 
below some multiple of the neutron-to-proton mass difference \citep{Qian.Woosley:1996}.
Without flavor oscillations, $Y_{e,{\rm fo}}>0.5$ and the NDW is proton-rich for both models
e8.8 and s27. As fast flavor oscillations increase the effective luminosity and
average energy of $\nu_e$ more than those of $\bar\nu_e$, they drive the NDW more proton-rich
(see Fig.~\ref{fig:dynamics}). Because the difference in luminosity and average energy between
$\nu_x$ and $\nu_e$ for model s27 is larger than that for model e8.8, the increase of $Y_{e,{\rm fo}}$
due to fast flavor oscillations for the former is larger than that for the latter. In addition,
for model e8.8, the luminosities and average energies of all neutrino 
species become closer with time, so the effect of fast flavor oscillations on $Y_{e,{\rm fo}}$ is 
greatly reduced at late times. The maximum values of $Y_{e,{\rm fo}}$ for different models of
the NDW are compared in Table~\ref{tab:wind_properties}.

Because the mass loss rate, $\tau_{\rm dyn}$, entropy, and $Y_{e,{\rm fo}}$ are set by
neutrino reactions well inside the wind termination point, the effects of fast flavor
oscillations on the above quantities are the same for model s27 with either $r_{\rm wt}=1000$ 
or 500 km. The main differences between these two cases are in the evolution of temperature
and density for $r>500$~km (see Fig.~\ref{fig:traj_examples}). The much slower expansion at later
times for the NDW with $r_{\rm wt}=500$~km amplifies the effects of $Y_{e,{\rm fo}}$ 
on nucleosynthesis by allowing more neutron production and subsequent proton capture 
(see \S\ref{sec:nucleo}).

\section{Effects of Fast Flavor Oscillations on Nucleosynthesis}
\label{sec:nucleo}

Proton-rich NDWs \citep{Martinez.Fischer.ea:2014,Arcones.Thielemann.ea:2013} 
are considered an important site for the $\nu p$ process
\citep{Bliss.Arcones.ea:2018,Fischer.Martinez-Pinedo.ea:2012,Froehlich.Martinez-Pinedo.ea:2006},
which can produce the light $p$-nuclei under suitable conditions, thereby affecting
their chemical evolution \citep{Travaglio.Rauscher.ea:2018}.
In a hot proton-rich environment, competition between $(p,\gamma)$ and $(\gamma,p)$ reactions
results in waiting-point nuclei that dominate the abundance in the associated isotopic chain,
and their $\beta$ decays control the progress to heavier nuclei.
In the $\nu p$ process, however, slow $\beta$ decays are overcome by $(n,p)$ reactions, for which
neutrons are provided by $\bar\nu_e$ absorption on protons \citep{Froehlich.Martinez-Pinedo.ea:2006}. 
For efficient production of heavy nuclei, neutron production should take place at
$\sim 1$--3~GK so that the temperature is sufficiently
high to overcome the increasingly high Coulomb barrier for further proton capture but low enough 
to avoid strong photo-dissociation. 

\subsection{Effectiveness of the $\nu p$ process}
The effectiveness of the $\nu p$ process is mainly determined by the proton-to-seed ratio
$Y_p/Y_h$ and the net neutron abundance provided by $\bar\nu_e$ absorption on protons.
The ratio $Y_p/Y_h$, where $Y_h$ is the number fraction of nuclei heavier than $^4$He,
specifies the availability of protons for neutron production and further proton capture
subsequent to $(n,p)$ reactions. Following \citet{Pruet.Hoffman.ea:2006}, we define
\begin{equation}
\label{eq:delta_n}
    \Delta_n=\left(\frac{Y_p}{Y_h} \right)_{T=3\,\mathrm{GK}} \times \int_{T<3\,\mathrm{GK}} \lambda_{\bar{\nu}_e p}\, dt,
\end{equation}
which measures the overall effectiveness of the $\nu p$ process. 
For a high $\Delta_n$, the NDW should have high entropy and fast expansion down to
$T\sim 3\,\mathrm{GK}$ to ensure a high $(Y_p/Y_h)_{T=3\,\mathrm{GK}}$, as well as
slow expansion at lower temperatures to ensure sufficient neutrino exposure for
neutron production. Slow expansion at $T\sim 1$--3~GK also facilitates further proton 
capture to make heavier nuclei.

\begin{figure}
    \centering
    \includegraphics[width=\linewidth]{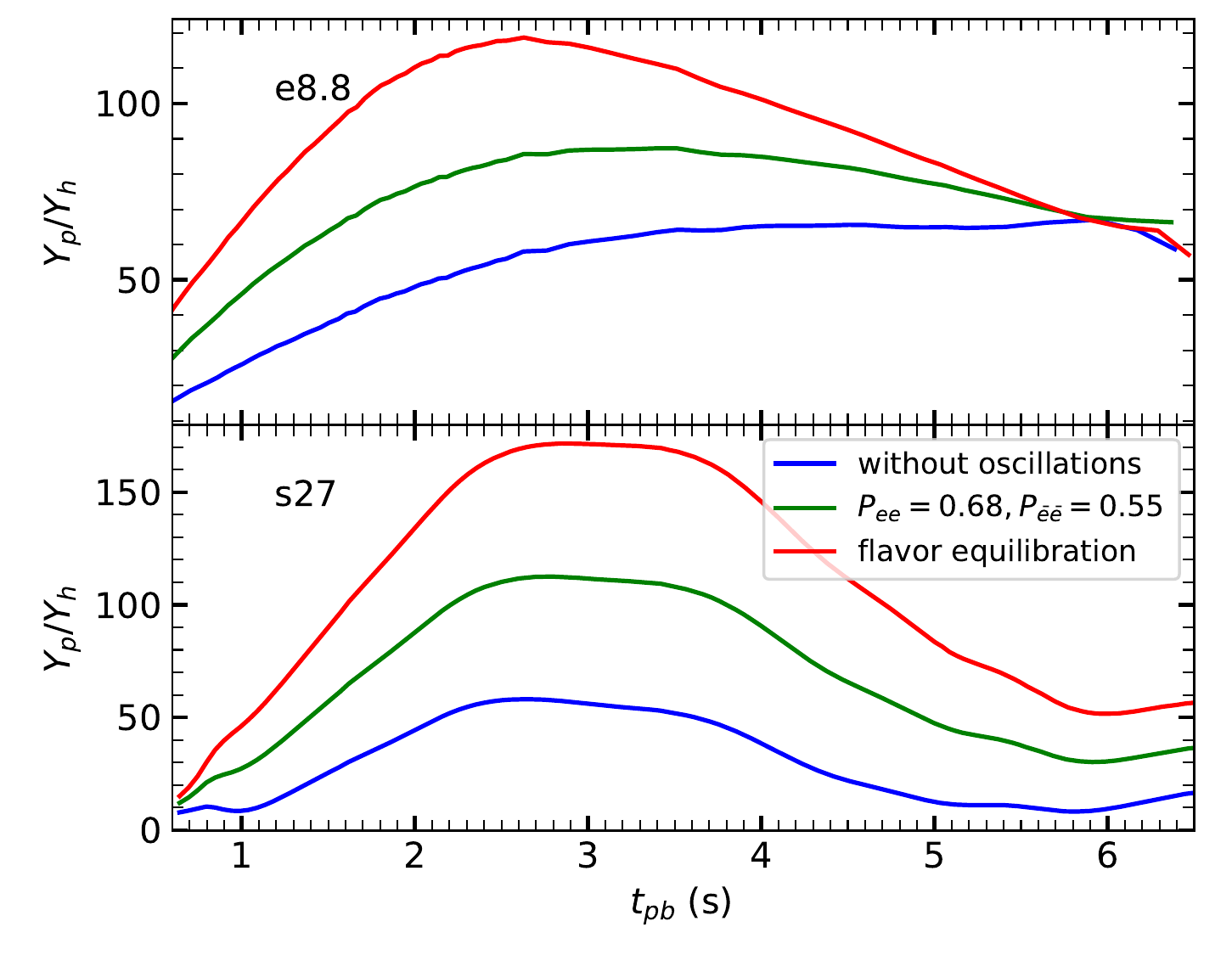}
    \caption{Proton-to-seed ratio $Y_p/Y_h$ at $T=3\,\mathrm{GK}$ for models e8.8 and s27
    with $r_{\rm wt}=1000$ km in the cases of no flavor oscillations (blue curves), 
    intermediate flavor conversion (green curves),
    and complete flavor equilibration (red curves).}
    \label{fig:p2s_e88}
\end{figure}

\begin{figure*}[!ht]
    \centering
    \includegraphics[width=\linewidth]{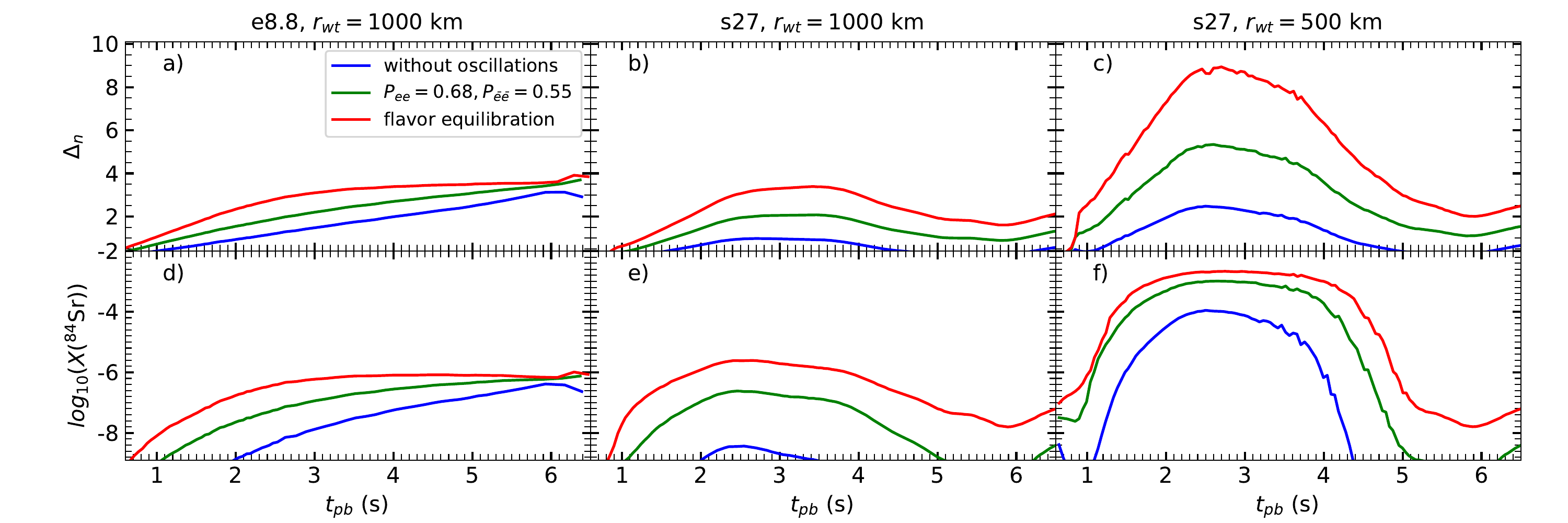}
	\caption{Comparison of $\Delta_n$ and $^{84}$Sr mass fraction for different NDW models
	in the cases of no flavor oscillations (blue curves), 
    intermediate flavor conversion (green curves),
    and complete flavor equilibration (red curves).}
    \label{fig:delta_n}
\end{figure*}

Figure~\ref{fig:p2s_e88} shows that fast flavor oscillations can increase
$(Y_p/Y_h)_{T=3\,{\rm GK}}$ by up to a factor $\sim 3$ ($\sim 2$) for model s27 (e8.8) 
with $r_{\rm wt}=1000$~km. Qualitatively, these effects are similar to those
on $Y_{e,\mathrm{fo}}$ shown in Fig.~\ref{fig:dynamics}. With the additional increase
of $\lambda_{\bar{\nu}_ep}$, fast flavor oscillations also increase $\Delta_n$
(see Fig, \ref{fig:delta_n}). While these effects are quite noticeable, $\Delta_n$ does not 
exceed 4 even with complete flavor equilibration for models e8.8 and s27 with 
$r_{\rm wt}=1000$~km (see Table~\ref{tab:wind_properties}). In contrast, for model s27 
with $r_{\rm wt}=500$~km, while $(Y_p/Y_h)_{T=3\,{\rm GK}}$ is the same as for 
$r_{\rm wt}=1000$~km, $\Delta_n$ is increased to up to $\sim 9$ with complete flavor 
equilibration because the duration corresponding to $T\sim1$--3~GK is much longer 
(see Fig.~\ref{fig:traj_examples}). The light $p$-nucleus $^{84}$Sr is the heaviest 
isotope noticeably produced in our models. Its mass fraction produced by each tracer 
is shown in Fig.~\ref{fig:delta_n} and clearly follows the trend of $\Delta_n$.

\subsection{Results on nucleosynthesis}
We calculate the nucleosynthesis for each tracer with the reaction network
used in \citet{Sieverding.Martinez.ea:2018}. Thermonuclear reaction rates are taken
from the ReaclibV2.2 library \citep{Cyburt.Amthor.ea:2010}, and $\beta$-decay
rates are taken from the NUBASE compilation of experimental
values \citep{Audi:2017} when available and from \citet{Moeller.Pfeiffer.ea:2003} otherwise. 
The only neutrino reactions included are those in Eq.~(\ref{eq:nupn})
with the same rates used for modelling the NDW. Beyond the wind termination point, 
we use, for example, 
\begin{equation}
    \lambda_{\bar\nu_ep}(t)=\lambda_{\bar\nu_ep}(t_{\rm wt})\left[\frac{r_{\rm wt}}{r(t)}\right]^2
    \exp\left(-\frac{t-t_{\rm wt}}{\tau_\nu}\right),
\end{equation}
where the exponential factor with $\tau_\nu=3$~s approximately accounts for the decrease of 
neutrino luminosities with time. Using the neutrino rates, which
include corrections for weak magnetism and nucleon recoil (see \citealt{Xiong.Wu.ea:2019} 
for details), we follow the evolution of $Y_e$ for a tracer taking into account
the change of the detailed nuclear composition. We find very good agreement with the
$Y_e$ calculated by the NDW models during the time when dynamic and thermodynamic
properties of the NDW are being determined. Apart from this check of consistency,
including the neutrino reactions, especially $\bar\nu_e$ absorption on protons, is 
required to follow the evolution of the neutron abundance for the $\nu p$ process.

For convenience, we label a tracer with the time $t_{\rm wt}$ at which it reaches 
the wind-termination radius $r_{\mathrm{wt}}$. Our reaction network calculations 
give the final mass fraction $X(Z,A,t_{\rm wt})$ for a nucleus 
with charge number $Z$ and mass number $A$ for the tracer. Each tracer is also
associated with a mass loss rate $\dot{M}(t_{\rm wt})$.
So the integrated yield of a nucleus is
\begin{equation}
\label{eq:integrated_yield}
 y(Z,A)=\int X(Z,A,t_{\rm wt})\,\dot{M}(t_{\rm wt}) \, dt_{\rm wt},
\end{equation}
which is evaluated with 72 and 184 tracers for models e8.8 and s27, respectively.

\begin{figure}
    \centering
    \includegraphics[width=\linewidth]{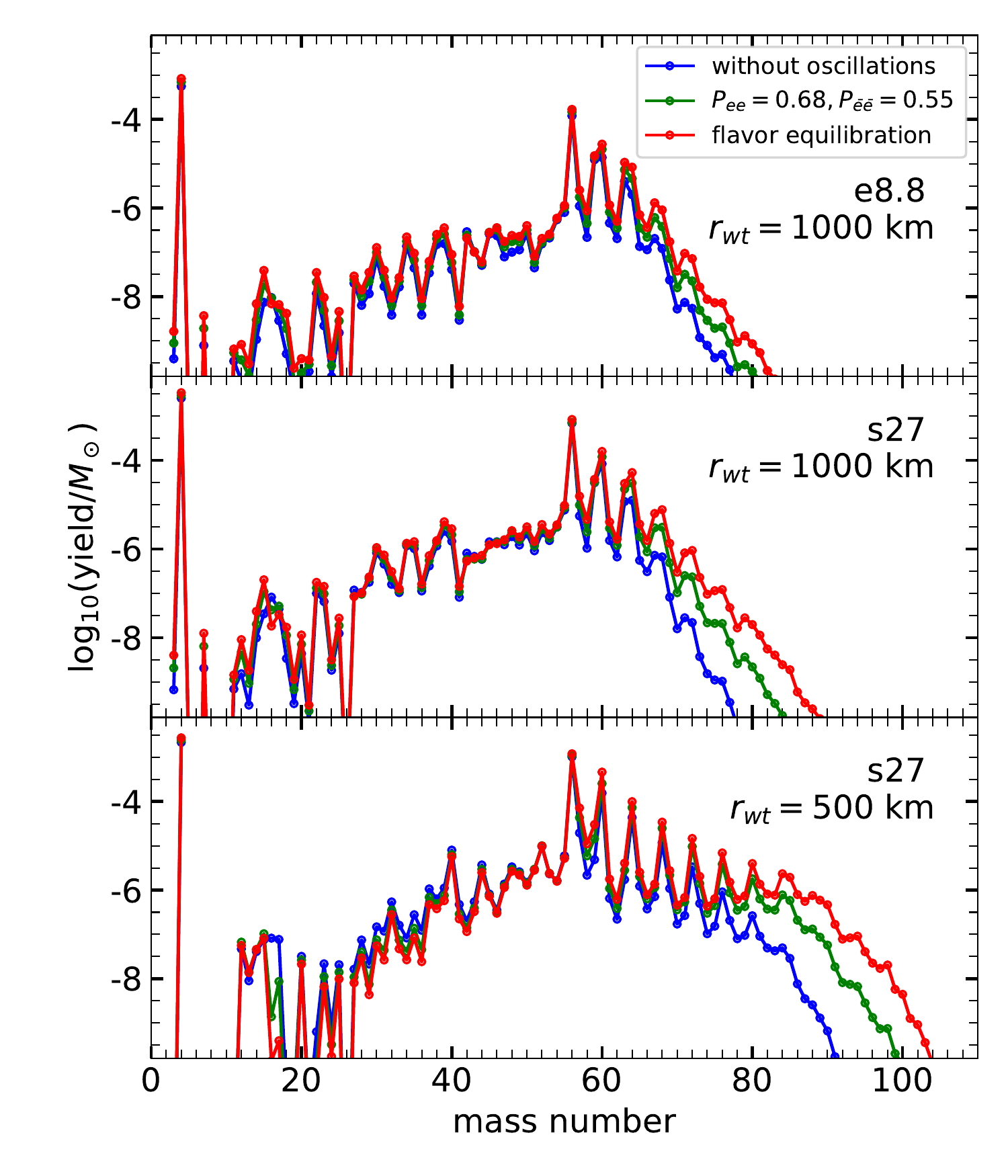}
	\caption{Comparison of nucleosynthesis yields for different NDW models
	in the cases of no flavor oscillations (blue curves), 
    intermediate flavor conversion (green curves),
    and complete flavor equilibration (red curves).}
    \label{fig:yields}
\end{figure}

The integrated yields are shown in Fig.~\ref{fig:yields} for different NDW models
in the cases of no flavor oscillations, intermediate flavor conversion, and
complete flavor equilibration. Yields of $^{64}$Zn and the light $p$-nuclei
$^{74}$Se, $^{78}$Kr, $^{84}$Sr, $^{92}$Mo, and $^{96}$Ru are also given in
Table~\ref{tab:yields}. The integrated yields contain large contributions 
from those NDWs whose conditions are not favorable to the $\nu p$ process
but only give rise to an $\alpha$-rich freeze-out.
The resulting yield patterns are very similar for models e8.8 and s27 with
$r_{\rm wt}=1000$~km in all the cases, with the mass fraction of
$^{4}$He being the largest followed by that of $^{56}$Fe. There are also
peaks at $^{60}$Ni with a closed shell of $Z=28$ protons and at $^{64}$Zn 
resulting from the decay of $^{64}$Ge. Note that $^{64}$Ge is the first 
important waiting-point nucleus for proton-rich nucleosynthesis 
\citep{Pruet.Woosley.ea:2005}. Its slow $\beta$-decay, which effectively
blocks production of heavier nuclei, is overcome by the $(n,p)$ reaction
following neutron production by $\bar\nu_e$ absorption on protons 
in the $\nu p$ process.

\begin{table*}
	\centering
	\caption{Yields (in $M_\odot$) for $^{64}$Zn and selected light $p$-nuclei for different NDW models [$X(Y)$ denotes $X\times 10^Y$].}
\begin{tabular}{ll|cccccc}
	\hline \hline
	Model &  Oscillations                       & $^{64}$Zn              & $^{74}$Se               & $^{78}$Kr               & $^{84}$Sr      & $^{92}$Mo  & $^{96}$Ru \\ \hline
e8.8, $r_{\rm wt}=1000$ km &   without oscillations              & $ 2.00(-6) $ &    $ 7.88(-10) $ &   $ 5.90(-11)  $ &  $ 3.55(-12)  $ &$   8.77(-15) $ &$  3.53(-16)$ \\   
e8.8, $r_{\rm wt}=1000$ km &   $P_{ee}=0.68, P_{\bar{e}\bar{e}}=0.55$  & $ 4.62(-6) $ &    $ 2.90(-9) $ &   $  2.58(-10) $ &  $ 1.91(-11)  $ &$   6.05(-14) $ &$  2.64(-15)$ \\ 
e8.8, $r_{\rm wt}=1000$ km &   flavor equilibration                & $ 8.33(-6) $ &    $ 8.68(-9)  $ &   $  9.39(-10) $ &  $ 9.49(-11)  $ &$   4.64(-13)  $ &$  2.42(-14)$ \\ \hline  
s27, $r_{\rm wt}=1000$ km  &   without oscillations              & $ 1.25(-5) $ &    $ 1.53(-9)  $ &   $  9.59(-11) $ &  $ 1.77(-12)  $ &$   7.99(-16)  $ &$  1.46(-17)$ \\  
s27, $r_{\rm wt}=1000$ km  &   $P_{ee}=0.68, P_{\bar{e}\bar{e}}=0.55$  & $ 3.08(-5) $ &    $ 2.16(-8)  $ &   $  2.56(-9)  $ &  $  1.74(-10) $ &$   4.42(-13)  $ &$  1.22(-14)$ \\  
s27, $r_{\rm wt}=1000$ km  &   flavor equilibration                & $ 5.33(-5) $ &    $ 9.57(-8)  $ &   $  1.64(-8)  $ &  $  2.44(-9)  $ &$    1.84(-11) $ &$  9.27(-13)$ \\ \hline  
s27, $r_{\rm wt}=500$ km  &   without oscillations              & $ 4.34(-5) $ &    $ 1.03(-7)  $ &   $  7.98(-8)  $ &  $  4.95(-8)  $ &$    6.70(-11) $ &$  3.54(-12)$ \\          
s27, $r_{\rm wt}=500$ km  &   $P_{ee}=0.68, P_{\bar{e}\bar{e}}=0.55$  & $ 7.27(-5) $ &    $ 3.00(-7)  $ &   $  3.51(-7)  $ &  $  7.76(-7)  $ &$    8.16(-9)  $ &$  1.33(-9) $ \\         
s27, $r_{\rm wt}=500$ km  &  flavor equilibration               & $ 9.92(-5) $ &    $ 4.34(-7)  $ &   $  6.14(-7)  $ &  $  2.34(-6)  $ &$    7.83(-8)  $ &$  2.24(-8) $ \\ \hline   
\end{tabular}
\label{tab:yields}
\end{table*}

For model e8.8, the yield exceeds $\sim 10^{-7}\,M_\odot$ for nuclei between $^{56}$Fe and $A=68$  
but drops steeply for heavier nuclei due to the inefficiency of the $\nu p$ process. 
Fast flavor oscillations increase the yields of nuclei heavier than $^{56}$Fe. In particular, the 
yield of $^{64}$Zn is increased by a factor of $\sim 4$ with complete flavor equilibration 
(see Table \ref{tab:yields}). Such oscillations, however, do not extend
significantly the range of heavy nuclei with yields of $\gtrsim 10^{-7}\,M_\odot$.
Specifically, the yields of the light $p$-nuclei are still rather low although they are
dramatically increased with fast flavor oscillations.
For model s27 with the same $r_{\rm wt}=1000$~km, the mass loss rates are higher than those for
model e8.8 in general (see Fig.~\ref{fig:dynamics}) and the highest $\Delta_n$ values are also
reached when the mass loss rates are relatively high (see Fig.~\ref{fig:delta_n}). Consequently,
the yields of nuclei heavier than $^{56}$Fe are higher than those for model e8.8 and
the effects of fast flavor oscillations are also noticeably stronger, extending nuclei with
yields of $\gtrsim 10^{-7}\,M_\odot$ up to the light $p$-nucleus $^{74}$Se (see Fig.~\ref{fig:yields}
and Table~\ref{tab:yields}).

For model s27 with $r_{\rm wt}=500$~km, the yield exceeds $\sim 10^{-7}\,M_\odot$ for heavy nuclei 
up to $A=80$ even without flavor oscillations, reflecting the corresponding relatively large $\Delta_n$
values (see Fig.~\ref{fig:delta_n}). With complete flavor equilibration, the range of nuclei with
yields of $\gtrsim 10^{-7}\,M_\odot$ is extended up to $^{90}$Zr and $^{91}$Zr, and even
the light $p$-nuclei $^{92}$Mo and $^{96}$Ru are produced with substantial yields of
$7.83\times 10^{-8}$ and $2.24\times 10^{-8}\,M_\odot$, respectively (see Table~\ref{tab:yields}).
Fast flavor oscillations also reduce the yields of nuclei with $A<60$ while increasing those
of heavier ones, clearly indicating the enhancement of the $\nu p$ process by such oscillations
(see Fig.~\ref{fig:yields}). 

\section{Implications}
\label{sec:implications}
In this section, we discuss the implications of our 
NDW models with fast flavor oscillations for abundances 
in metal-poor stars, Galactic chemical 
evolution in general, and isotopic anomalies in meteorites.

\subsection{Abundances in metal-poor stars}
Metal-poor stars were formed out of the interstellar medium (ISM) at very early times  
and their surface abundances reflect chemical enrichment by only a small number of 
nucleosynthetic events prior to their formation \citep[e.g.,][]{Frebel:2018}.
Consequently, the abundances in such a star may show some features of the 
nucleosynthesis of a CCSN at low metallicities. Because $^{64}$Zn is a major product 
of NDWs, which most likely remains true of low-metallicity CCSNe, we estimate 
[Zn/Fe]~$=\log({\rm Zn/Fe})-\log({\rm Zn/Fe})_\odot$ using the Zn yields
from our NDW models along with the Fe yields from the associated CCSNe, and compare
the results with the data on metal-poor stars. We include all the isotopes of
the same element in the results presented below.

For the CCSN associated with model e8.8, explosive nucleosynthesis produces
$\sim 0.002$--$0.004\,M_\odot$ of Fe \citep{2009ApJ...695..208W}, which greatly exceeds
the amount [$\sim(1$--$2)\times 10^{-4}\,M_\odot$] produced in the NDWs with or without
flavor oscillations. Assuming that the Zn and Fe yields apply at low 
metallicities, we expect [Zn/Fe]~$\sim -0.5$ to $-0.2$, $-0.1$ to 0.2, and 0.2 to 0.5
from our NDW models without flavor oscillations, with intermediate flavor conversion, and with 
complete flavor equilibration, respectively. The above results on [Zn/Fe] would correspond to
[Fe/H]~$\sim -3.7$ to $-2.4$ with the CCSN enriching $\sim 10^3$--$10^4\,M_\odot$ of ISM. 
Observations \citep{Cayrel.Depagne.ea:2004} show that stars
with [Fe/H]~$\sim -4.4$ to $-2.5$ have [Zn/Fe]~$\sim 0.1$--0.7. So it appears that
NDWs from CCSNe with progenitors and neutrino emission similar to those of model e8.8
can account for some of the data on [Zn/Fe] for metal-poor stars only when fast flavor oscillations
are included.

For low-metallicity CCSNe with progenitors and neutrino emission similar to those of model s27, 
we expect a much wider range of Fe yields due to the great uncertainties in the explosion
mechanism and the role of mixing and fallback. Assuming the Zn yields from our NDW models, 
we find that a nominal Fe yield of
$\sim 0.1\,M_\odot$ would give low values of [Zn/Fe]~$<-0.1$ for model s27 with
$r_{\rm wt}=1000$ or 500 km regardless of flavor oscillations. 
It is also plausible, however,
that the low-metallicity CCSNe of concern undergo weak explosion,
in which most of the He-core falls back onto the PNS (most likely resulting in the formation of a black hole) 
and only a small fraction of the inner nucleosynthesis products are mixed into the ejecta.
The transitional 
existence of the PNS would still allow a NDW to form and a fraction of this material to be mixed into 
the ejecta.
We explore such a scenario below to explain the high values of both [Zn/Fe] and [Sr/Fe] 
in the metal-poor star HE 1327--2326 \citep{Ezzeddine.Frebel.ea:2019}.

Because of the uncertainties in the Fe yield and the ejected fraction, we first seek to
explain [Sr/Zn]~$\sim 0\pm0.3$ for HE 1327--2326 \citep{Ezzeddine.Frebel.ea:2019}.
As our model s27 with $r_{\rm wt}=500$~km is more consistent with a weak explosion, we focus on
this case. Assuming that Sr and Zn made in the NDW are ejected with the same fraction, 
we obtain [Sr/Zn]~$=-1.4$, $-0.3$, and 0.1 without flavor oscillations, with intermediate
flavor conversion, and with complete flavor equilibration, respectively. Clearly, only
with fast flavor oscillations is it possible to explain the observed [Sr/Zn] by the NDW. Note that
while $^{64}$Zn is the dominant Zn isotope, the contribution from other Sr isotopes is comparable
to that from $^{84}$Sr for complete flavor equilibration. To explain the observed
[Zn/Fe]~$=0.80\pm0.25$, we note that the Zn and Fe from the NDW would give [Zn/Fe]~$=1.7$--1.8
with fast flavor oscillations, which represents an upper limit had there been no Fe contributed 
from explosive nucleosynthesis. The observed [Zn/Fe], as well as [Sr/Fe]~$={\rm [Sr/Zn]}+{\rm [Zn/Fe]}$,
can be accounted for if
the ejected Fe from explosive nucleosynthesis is $\sim 10$ times more than that
from the NDW. 

While looking for a CCSN model that matches the above constraints is beyond 
the scope of this study, we can use the observed value of 
[Fe/H]~$=-5.2$ for HE 1327--2326 \citep{Ezzeddine.Frebel.ea:2019}
to estimate the fraction of NDW material that needs to be included in the ejecta.
To explain this [Fe/H],
$\sim 10^{-5}$--$10^{-4}\,M_\odot$ of Fe is needed to mix with $\sim 10^3$--$10^4\,M_\odot$ of ISM.
As only $\sim 10\%$ of this Fe is from the NDW but $\sim 10^{-3}\,M_\odot$ of Fe is produced there,
we estimate $\sim (0.1$--1)\% of the NDW material is mixed into the ejecta.
We consider that the above scenario of a low-metallicty CCSN with mixing and fallback, including
the NDW contributions with fast flavor oscillations, may explain the
high values of both [Zn/Fe] and [Sr/Fe] for HE 1327--2326. We refer readers to
\citet{Woosley.Heger.ea:2010} and \citet{Ezzeddine.Frebel.ea:2019} for detailed discussions
of CCSN models that may fit the overall abundance pattern of this star.

\subsection{Galactic chemical evolution}
\begin{figure*}[!ht]
    \centering
    \includegraphics[width=\linewidth]{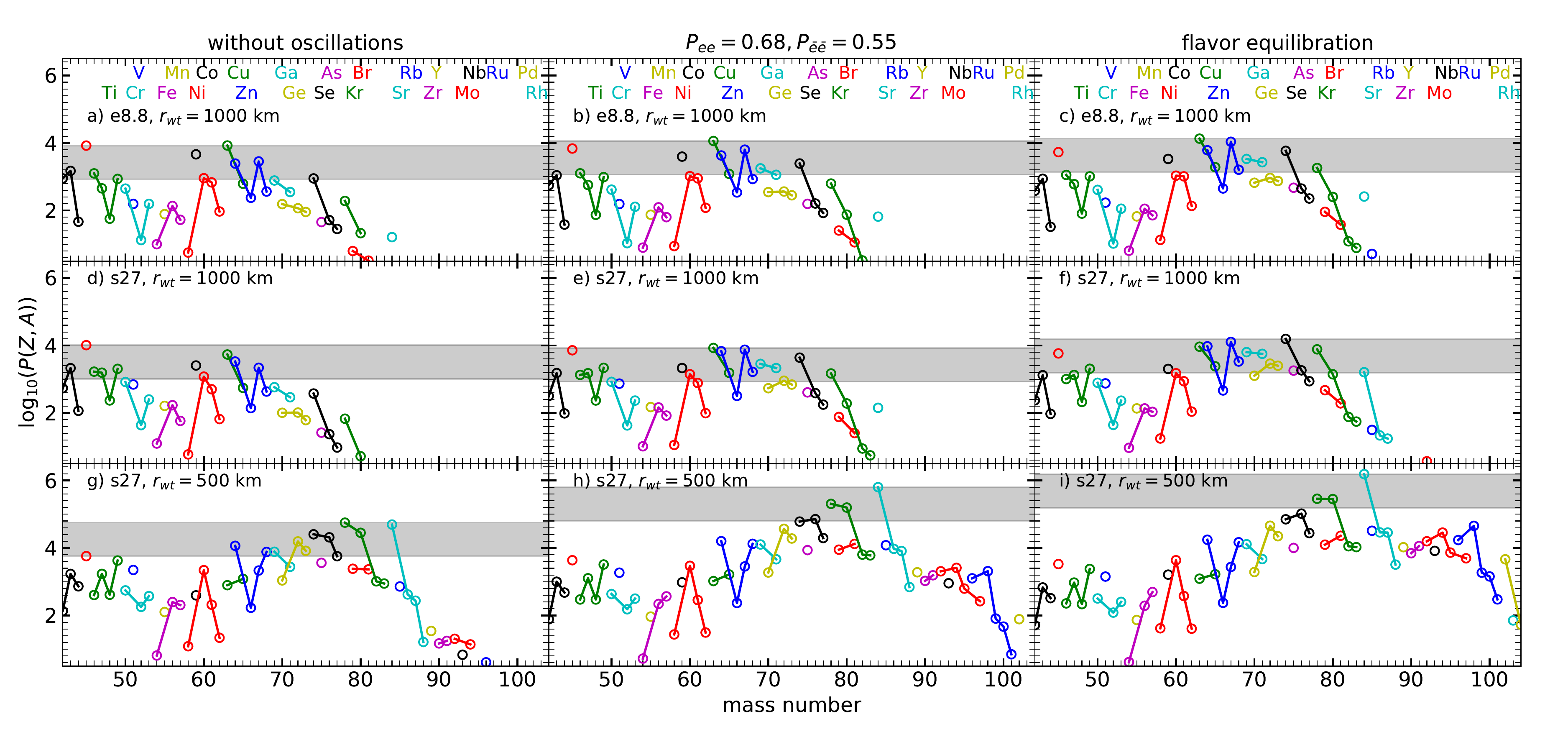}
    \caption{Production factors $P(Z,A)$ for all the NDW models. Isotopes of 
    the same element are connected by line segments. 
    Nuclei in the grey bands have production factors exceeding 0.1 times
    the corresponding maximum value.}
    \label{fig:abu_compare}
\end{figure*}

We expect that the yields from our NDW models apply generally to
NDWs in CCSNe with progenitors and neutrino emission similar to those
considered here. Under this expectation, we can estimate the dominant
contributions from such NDWs to the present Galactic inventory of
nuclei. Taking the solar composition \citep{Asplund.Grevesse.ea:2009}
as representative of this inventory, we can identify the significant
nuclei contributed from a NDW model by considering the production
factor $P(Z,A)=X(Z,A)/X_\odot(Z,A)$, where $X(Z,A)$ is the mass fraction
of the nucleus with charge number $Z$ and mass number $A$ produced in 
the model. Figure~\ref{fig:abu_compare} shows $P(Z,A)$ for all our
models. Nuclei within the grey bands have production factors
exceeding 0.1 times the maximum value $P_{\rm max}$, and therefore,
their production by NDWs can potentially make significant to
dominant contributions to their present Galactic inventory.
The actual NDW contributions, however, also depend on the net mass
loss and the frequency of occurrences for the relevant source.

We first discuss the nuclei in the grey bands in Fig.~\ref{fig:abu_compare}.
For model e8.8 without flavor oscillations, $P_{\rm max}=8.4\times10^3$ corresponds to
$^{63}$Cu, a result of proton-rich freeze-out from NSE
\citep[e.g.,][]{Frohlich.Hauser.ea:2006}. Other nuclei in the band
include $^{67}$Zn, $^{64}$Zn, and the light $p$-nucleus $^{74}$Se.
With fast flavor oscillations, $P_{\rm max}$ still corresponds to
$^{63}$Cu, but its value is increased by a factor of 1.4--1.6.
The production factors of $^{67}$Zn, $^{64}$Zn, and $^{74}$Se are 
increased by a factor of 2.2--3.8, 1.7--2.5, and 2.8--6.6, respectively.
In addition, the light $p$-nucleus $^{78}$Kr is added to the grey band
for complete flavor equilibration. The general effects of fast flavor
oscillations to increase production factors and extend the range of
significant production to heavier nuclei also apply to model s27.
For $r_{\rm wt}=1000$~km, $P_{\rm max}=1.0\times10^4$, $8.4\times10^3$, 
and $1.6\times10^4$ corresponds to $^{45}$Sc without flavor oscillations, 
$^{63}$Cu with intermediate flavor conversion, and $^{74}$Se 
with complete flavor equilibration,
respectively. The nuclei $^{67}$Zn, $^{64}$Zn, and $^{74}$Se
have production factors of $\sim(0.5$--$1)P_{\rm max}$
with fast flavor oscillations. In addition, $^{78}$Kr is added to the
grey band for intermediate flavor conversion and another
light $p$-nucleus $^{84}$Sr is added for complete flavor equilibration.
For $r_{\rm wt}=500$~km, $^{78}$Kr and $^{84}$Sr have the largest
production factors with or without flavor oscillations. Specifically,
$P(^{78}{\rm Kr})=5.6\times10^4$, $2.0\times10^5$, and $2.9\times10^5$
[$P(^{84}{\rm Sr})=4.9\times10^4$, $6.3\times10^5$, and $1.5\times10^6$] without
flavor oscillations, with intermediate flavor conversion, and with 
complete flavor equilibration, respectively.

Assuming that each NDW model represents a CCSN source 
occurring in a total mass $M_{\rm ISM}\sim 10^{11}\,M_\odot$ of ISM with an average frequency
$f_{\rm NDW}$ over the Galactic history of duration $t_G\sim 10^{10}$~yr, we can
estimate the fraction $f(Z,A)$ contributed by this source to
the Galactic inventory of a nucleus with charge number $Z$ and mass number $A$:
\begin{align}
    f(Z,A)&=\frac{X(Z,A)M_{\rm NDW}f_{\rm NDW}t_G}{X_\odot(Z,A)M_{\rm ISM}}\nonumber\\
    &\sim\frac{P(Z,A)}{3\times 10^5}\left(\frac{M_{\rm NDW}}{10^{-3}\,M_\odot}\right)
    \left[\frac{f_{\rm NDW}}{(30\ {\rm yr})^{-1}}\right],
\end{align}
where $M_{\rm NDW}$ is the net mass loss of the NDW (see Table~\ref{tab:wind_properties}).
For reasonable values of $f_{\rm NDW}\lesssim(30\ {\rm yr})^{-1}$,
model e8.8 could contribute at most $\sim 2\%$, $\sim4\%$, and $\sim6\%$ of the Galactic 
inventory of $^{63}$Cu without flavor oscillations, with intermediate flavor conversion, and
with complete flavor equilibration, respectively. Because $^{63}$Cu has the largest production
factor, we conclude that these NDWs cannot make major contributions to the Galactic inventory
of any nucleus. Model s27 with $r_{\rm wt}=1000$~km could make more significant contributions
to the Galactic inventory of the dominant products: up to $\sim 12\%$ for $^{45}$Sc without
flavor oscillations, $\sim12\%$ for $^{63}$Cu with intermediate flavor conversion, and 
$\sim28\%$ for $^{74}$Se with complete flavor equilibration. Nevertheless, no major
contributions to any nucleus are expected from these NDWs, either.

Model s27 with $r_{\rm wt}=500$~km is the most interesting. Without flavor oscillations,
it could contribute up to $\sim68\%$ of the $^{78}$Kr and $\sim 60\%$ of the $^{84}$Sr
in the Galactic inventory. Because the production factors are increased significantly 
by fast flavor oscillations, we must limit the frequency of occurrences of the corresponding
NDWs to avoid overproducing any nucleus, especially the dominant product $^{84}$Sr. 
Specifically, for the model under discussion,
NDWs with intermediate flavor conversion (complete flavor equilibration) could occur only 
with an average frequency of $\lesssim (280\,{\rm yr})^{-1}$ 
[\,$\lesssim (850\,{\rm yr})^{-1}$], corresponding to $\lesssim 11\%$ ($\lesssim 3.5\%$)
of the CCSNe.

The $\gamma$ process can also produce $p$-nuclei during the evolution of
a massive star and the explosive burning of CCSN ejecta. The yield of $^{84}$Sr by this process
varies greatly from $\sim 10^{-9}$ to $\sim 10^{-6}\,M_\odot$ \citep[e.g.][]{Pignatari.Herwig.ea:2016}.
For model s27 with $r_{\rm wt}=1000$~km, the $^{84}$Sr yield from NDWs reaches the lower end of this
range only with complete flavor equilibration (see Table~\ref{tab:yields}). 
For model s27 with $r_{\rm wt}=500$~km, however,
the $^{84}$Sr yield from NDWs is already significant even without flavor oscillations and reaches the 
higher end of the above range with fast flavor oscillations. As we have shown above, if such
NDWs with fast flavor oscillations were to occur in a relatively small fraction of CCSNe, they could 
dominate the Galactic inventory of $^{84}$Sr. Of course, more studies are needed to quantify
the actual contributions from these NDWs and the $\gamma$ process.

\subsection{Isotopic anomalies in meteorites}
The widely-varying production factors for each NDW model shown in Fig.~\ref{fig:abu_compare}
mean that the production patterns for these NDWs differ greatly from the solar pattern.
Consider CCSNe occurring at times close to the formation of the solar system. If some of the 
associated NDW material could condense into grains that would be incorporated into meteorites,
then isotopic anomalies characteristic of NDW nucleosynthesis would be found in such meteorites. 
Specifically, there would likely be excesses of those nuclei in the grey bands shown in 
Fig.~\ref{fig:abu_compare}. Note that even when a NDW source is not a major contributor to
the Galactic inventory, it still can produce isotopic anomalies in meteorites so long as
the relevant grains could be formed and incorporated into meteorites.

An excess of $^{84}$Sr was found in Ca-Al-rich inclusions (CAIs) in the Allende meteorite
\citep[e.g.,][]{2013PNAS..11017241B,2019GeCoA.261..145B}. As shown in Fig.~\ref{fig:abu_compare},
this nucleus is the dominant product of NDWs for model s27 with $r_{\rm wt}=500$~km and
with fast flavor oscillations while its isotopes are greatly suppressed. Therefore,
such NDWs could be the source producing the observed excess of $^{84}$Sr.
Interestingly, excesses of $^{92,94,95,97,100}$Mo were also found in the same Allende
CAIs \citep{2013PNAS..11017241B}. While our NDW models cannot account for such excesses,
we note that neutron-rich NDWs, possibly from NSMs, could be a source for 
these Mo anomalies \citep{Bliss.Arcones.ea:2018}. 

\section{Discussion and Conclusions}
\label{sec:conclusions}
Using the unoscillated neutrino emission characteristics from two CCSN simulations
(models e8.8 and s27) representative of relevant progenitors at the lower and higher mass end,
we have studied the potential effects of fast flavor oscillations on NDWs and their nucleosynthesis.
We find that such oscillations can increase the total mass loss by factors up to $\sim 1.5$--1.7
and lead to significantly more proton-rich conditions. The latter effect can greatly enhance the 
production of $^{64}$Zn and the so-called light $p$-nuclei $^{74}$Se, $^{78}$Kr, and $^{84}$Sr.

Were fast flavor oscillations to occur close to the PNS surface as described here, they
would have important implications for abundances in metal-poor stars, Galactic chemical evolution
in general, and isotopic anomalies in meteorites (see \S\ref{sec:implications}). 
When such oscillations are included, the Zn from NDWs in model e8.8 could readily explain some of 
the data on [Zn/Fe] for metal-poor stars \citep{Cayrel.Depagne.ea:2004}. 
In addition, the Zn and Sr from NDWs in model s27 with $r_{\rm wt}=500$~km could explain the high 
values of both [Zn/Fe] and [Sr/Fe] for HE 1327--2326 \citep{Ezzeddine.Frebel.ea:2019} in the context 
of mixing and fallback for a low-metallicity CCSN.
While NDWs in models e8.8 and s27 with $r_{\rm wt}=1000$~km do not make major contributions
to the present Galactic inventory even with fast flavor oscillations, their clearly non-solar
production patterns might result in isotopic anomalies if some NDW material could
condense into grains that would be incorporated into meteorites. The observed excess of
$^{84}$Sr in the Allende CAIs \citep[e.g.,][]{2013PNAS..11017241B,2019GeCoA.261..145B}
might be explained by a source like the NDWs in model s27
with $r_{\rm wt}=500$~km and with fast flavor oscillations.
The dominant production of $^{84}$Sr by such a source, however, leads to a constraint that it
can represent only $\lesssim (3.5$--11)\% of CCSNe to avoid overproducing this nucleus.

\citet{Roberts.Woosley.ea:2010} performed a hydrodynamic study of NDWs based on
neutrino emission from CCSN simulations, including our model e8.8.
Allowing for the effects of fast flavor oscillations not included in their study, their results
are qualitatively similar to ours. There were also many parametric studies of NDW nucleosynthesis
\citep[e.g.,][]{Wanajo.Janka.ea:2011,Bliss.Arcones.ea:2018,Nishimura.Rauscher.ea:2019}. 
The parameterization can be tuned to provide more extreme values of the entropy, $Y_e$, 
and hence, $\Delta_n$, than what we find in our NDW models. When similar conditions
are considered, our results are consistent with those of e.g., \citet{Wanajo.Janka.ea:2011}.
The conditions in our NDW models are similar to those for the early proton-rich inner ejecta
heated by neutrinos in multi-dimensional CCSN simulations \citep{Wanajo.Mueller.ea:2018},
which result in characteristic production of $^{45}$Sc and $^{64}$Zn by
proton-rich freeze-out from NSE \citep{Frohlich.Hauser.ea:2006}. 
Recent 3D simulations by \citet{Burrows.Radice.ea:2020} also confirmed the predominant
proton-richness of the inner ejecta and reported a wind-like component.
We expect that the potential 
effects of fast flavor oscillations on the nucleosynthesis in such ejecta are 
qualitatively similar to those found here for the NDWs. The early neutron-rich inner
ejecta \citep[e.g.,][]{Eichler.Nakamura.ea:2018}, however, require a separate study.

While we have explored the potential effects of fast flavor oscillations on NDWs
and their nucleosynthesis, two important questions remain to be investigated for various
CCSN simulations: can such oscillations occur close to the PNS surface,
and if so, with what flavor conversion efficiency? Further efforts to address the above 
questions are crucial to support the important implications of fast flavor oscillations
for abundances in metal-poor stars, Galactic chemical evolution in general, and isotopic 
anomalies in meteorites as discussed here.

\begin{acknowledgments}
  We thank H.-T. Janka for providing access to the data on neutrino emission from the 
  CCSN simulations by his group.
  This work was supported in part by the US Department of Energy
  [DE-FG02-87ER40328 (UM)]. Some calculations were performed at the Minnesota 
  Supercomputing Institute. MS acknowledges support from the National Science Foundation, 
  Grant PHY-1630782, and the Heising-Simons Foundation, Grant 2017-228.
 \end{acknowledgments}

\end{document}